\documentclass[dvips]{article}
\usepackage{graphicx}

\title{HiRes Estimates and Limits for Neutrino Fluxes at the Highest Energies}
\author{K. Martens$^1$ (for the High Resolution Fly's Eye Collaboration)}

\begin{document}
\maketitle

The High Resolution Fly's Eye Experiment (HiRes) measures cosmic 
rays (CR) at the highest energies using the air fluorescence technique. 
As data taking on the Dugway Proving Grounds in Western Utah is finished, 
the HiRes data are relevant for cosmogenic neutrinos in two different ways. 
We first use our best fit to the measured HiRes CR spectrum together with a 
model of the extragalactic CR sources to derive the 
expected cosmogenic neutrino and gamma ray fluxes at Earth. 
We then use the HiRes data directly to set competitive experimental limits 
on the electron and tau neutrino fluxes at the highest energies. 


\mbox{}\\
$^1$University of Utah, Department of Physics \\
115 S. 1400 E., Salt Lake City, UT 84112-0830, USA

\section{Introduction}

Cosmogenic neutrinos are neutrinos generated by cosmic rays (CR) as they 
propagate 
over cosmological distances and lose energy in interactions on various photon
backgrounds. 
Excitation of the $\Delta^+$ resonance in protons moving through the Cosmic 
Microwave Background (CMB) will start for proton energies above 
$\sim 6 \times10^{19}$~eV, leading to the expectation that these protons will 
arrive at earth with energies diminished by this energy loss process. 
The resulting feature at the end of the CR power law spectrum is called the 
GZK cutoff after the authors who first pointed out this theoretical limit to 
the reach of a protonic CR spectrum: Greisen, Zatzepin, and Kuzmin \cite{bib:GZK}.
With the GZK feature evident in the HiRes monocular spectra \cite{bib:HR-GZK}, 
and with the Auger CR spectrum in good agreement with HiRes at the 
relevant energies, there now is a strong case for the existence of cosmogenic
neutrinos.

This interpretation of the CR spectrum hinges on the 
assumption that the extragalactic CRs at the highest energies are 
protonic in nature. While composition studies for CR do not reach up to the 
highest energies yet, HiRes has measured the composition up to 
2.5$\times$10$^{19}$, concluding that at the highest energies the CR indeed are 
mostly protons \cite{bib:Comp}. 

\section{The HiRes Detector}

The HiRes experiment consists of two detector sites located
12.6~km apart that each monitor the surrounding sky for fluorescence emission
from extensive air showers. On clear, moonless nights individual telescopes
continuously survey a patch of the sky that measures roughly 16~degree along
the horizon and 14~degree in elevation. 256 photomultiplier tubes per
telescope allow a pixelisation with a pixel size of 1 degree in the sky.
If all telescopes are taking data, both detectors achieve nearly full coverage 
in azimuth.
HiRes 1 (HR1) has 22 telescopes viewing the elevation band from 3~degree
above the horizon to 17~degree above the horizon. It is equipped with
sample-and-hold electronics that record the threshold crossing time for
each pixel and a charge integral over 3.6~$\mu$s.
HiRes 2 (HR2) has 42 mirrors that cover elevations from 3 to 33~degree.
It is equipped with flash-ADC electronics that trace the signal evolution for
every pixel with 100~ns resolution over 10~$\mu$s.

\section{Neutrino and Gamma Ray Fluxes Expected from the 
HiRes Monocular Spectra}

The HiRes detector was developed for stereoscopic observation of extensive 
air showers (EAS) by means of their fluorescence emission. The restriction to 
stereoscopically seen EAS limits the HiRes aperture in two important ways: 
At the highest energies detector lifetime is most important, and using older 
equipment HR1 was operational for almost 4 years before HR2 became 
operational to allow stereoscopic observation. At the lower energies the 
reduced fluorescence light output of the lower energy EAS forces us to 
rely on the HR2 monocular observations as HR2 has higher elevation coverage 
and can thus capture the shower maximum even for close-by EAS. 
For these reasons the monocular spectra of the two HiRes 
detectors will always be superior in statistics and, most important for 
spectrum fits, span a wider range of energies than the stereo spectrum. 

Building on work of Doug Bergman at Rutgers \cite{bib:Doug}, we calculate the 
neutrino and gamma ray spectra expected due to the propagation of protons from 
their hypothetical source population to Earth. The CR spectrum as measured by 
HiRes detectors in monocular mode is 
used to constrain the model of astrophysical proton sources. Protons emitted 
at the source undergo interactions with the CMB that excite the $\Delta^+$ 
resonance or produce $e^+e^-$ pairs and are cooled down by the expansion of 
the universe along their way. All these processes are taken into account in 
calculating what the proton spectrum will be at earth by adding the 
respective contributions from redshifts out to 4.0 in steps of 0.01. The 
sources are all assumed to have the same universal injection spectrum of the 
form $E^{-\gamma}$, and $\gamma$ together with a universal normalization 
constant for this injection spectrum are two of the three free parameters 
used in fitting the HiRes CR spectra to this source model. The last parameter 
is an exponent $m$ in the form $(1+z)^m$ that we assume for the evolution of 
this source population with redshift $z$. 10,000 protons are propagated 
through an evolving CMB for each of our shells of 0.01 thickness in $z$. 

As first pointed out by Berezinsky \cite{bib:Berezinsky}, this model 
reproduces well 
the features of the CR spectrum at the highest energies. The ankle in this 
context is excavated by the $e^+e^-$ energy loss mechanism and does not mark 
the transition from galactic to extragalactic sources. This latter transition 
is marked by the disappearance of Fe nuclei from the composition at end of the 
galactic CR spectrum. Above about $10^{18}$~eV this transition from galactic to 
extragalactic CRs is completed as signalled by a change of slope in 
our composition measurement. In our fits to the HiRes CR spectrum we add a 
galactic component of Fe nuclei according to the fractional composition of the 
measured spectrum as suggested by our composition measurements. 

\begin{figure}[h]
\begin{center}
\includegraphics [width=0.98\textwidth]{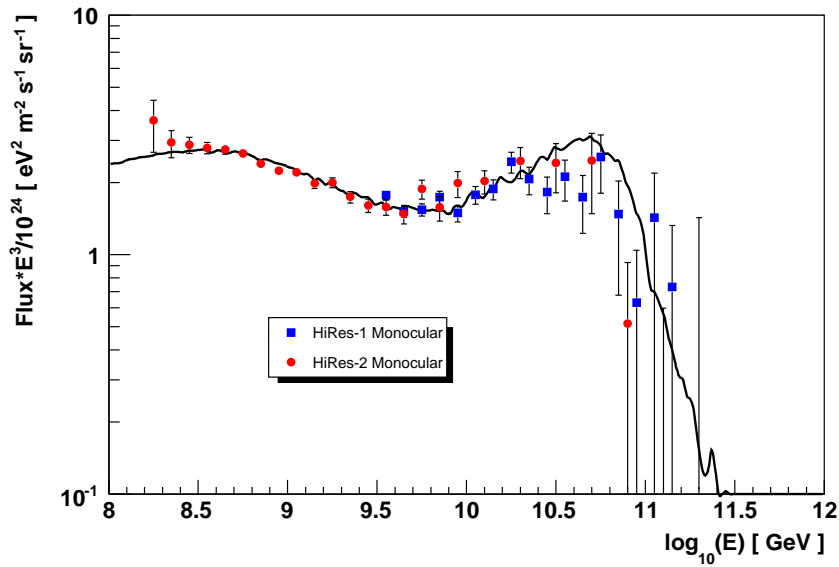}
\end{center}
\caption{
The best fit to the HiRes monocular spectra propagating protons from 
extragalactic sources up to redshifts of 4. A component of iron nuclei is 
added as derived from HiRes composition measurements. 
}
\label{fig:fit}
\end{figure}

The resulting best fit to the HiRes monocular CR spectrum is shown in Figure 
\ref{fig:fit}. It is obtained for $\gamma=2.42$ and $m=2.46$. In this fit the 
low energy region is most affected by changes in $\gamma$, whereas the high 
energy region reacts more strongly to changes in $m$. Future improvements to 
these fits will include a more realistic redshift evolution modelled after 
observations on suitable populations of Active Galactic Nuclei, the prime 
suspects in accelerating CR to the energies involved here. The spikes that 
appear at high energies in the fitted proton spectrum reflect our choice of 
stepsize in $z$. Future work will reduce this stepsize from its current 
value of 0.01 to arrive at a smoother distribution. 

\begin{figure}[h]
\begin{center}
\includegraphics [width=0.98\textwidth]{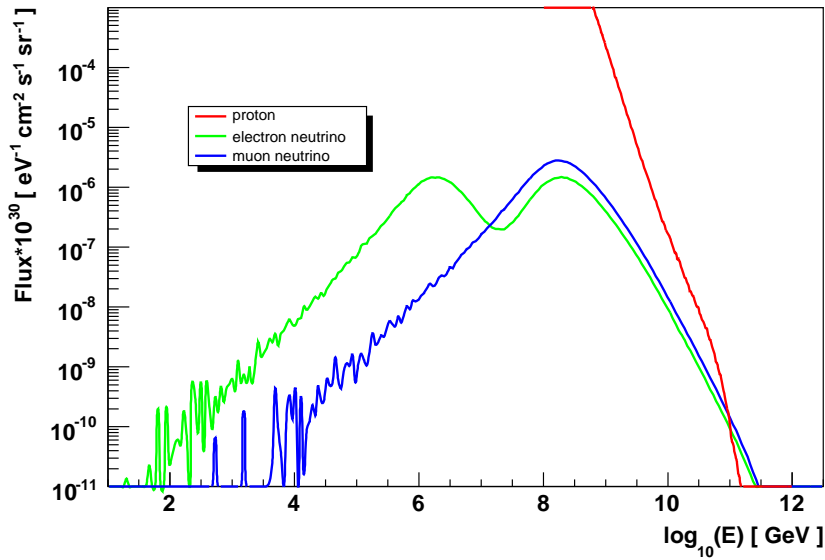}
\end{center}
\caption{
Fluxes of electron (green) and muon (blue) neutrinos. The neutrinos are shown 
with their flavors at origin but their energies properly redshifted to Earth. 
The red proton spectrum is the one from the best fit and shown for reference. 
}
\label{fig:neu}
\end{figure}

\begin{figure}[h]
\begin{center}
\includegraphics [width=0.98\textwidth]{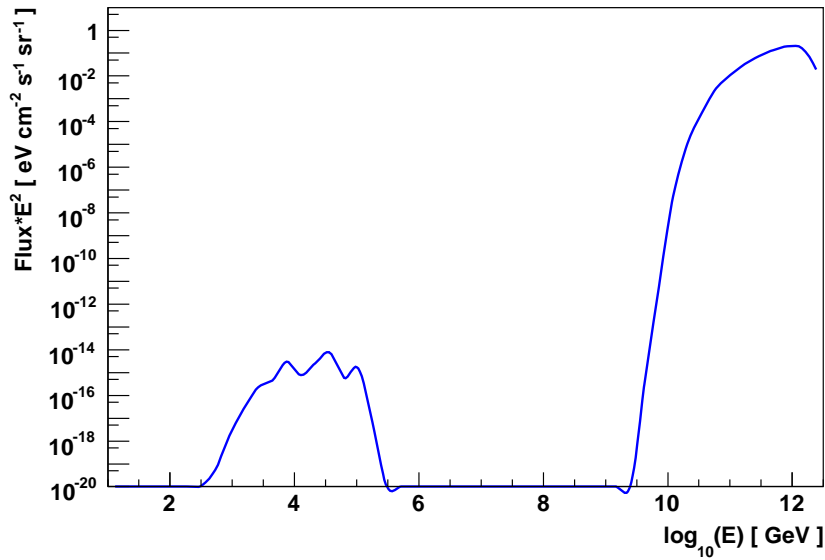}
\end{center}
\caption{
Gamma ray fluxes for their energies upon arrival at Earth. The wiggles in the 
low energy bump are due to statistical limitations. The fate of the $e^+$ and 
$e^-$ produced in the CMB interactions of the photons that were lost in 
propagation is not followed up on and resulting secondary photon fluxes do not 
contribute to this spectrum. 
}
\label{fig:gam}
\end{figure}

Figures \ref{fig:neu} and \ref{fig:gam} show the respective neutrino and 
gamma ray fluxes at the highest energies resulting from CMB interactions of 
the propagating protons. The ``low'' energy hump in the electron neutrino 
spectrum originates from the neutron decay occurring if the $\Delta^+$ 
resonance decays into $n + \pi^+$. The structure in the gamma ray spectrum 
is carved out by interactions with the CMB of the gamma rays themselves as 
they propagate from their respective points of origin to Earth. The spikiness 
in the gamma ray spectrum near 10~TeV is an artefact of limited statistics. 

\section{Neutrino Flux Measurements with HiRes}

Two different neutrino flux limits exist from the HiRes experiment. The 
first one exploits the fact that charged current tau neutrino interactions in 
the vast target mass of mountains and the Earth's crust produce tau leptons 
that can escape into the atmosphere, where their decay may then produce an 
air shower that is visible to the HiRes detector. The second limit exploits 
the Landau, Pomeranchuk, Migdal (LPM) effect for electron neutrino induced 
electromagnetic showers in the earth and in the atmosphere.

\section{Monte Carlo Simulation}

Our Monte Carlo (MC) simulation of tau neutrino interaction and tau propagation
and decay is based on the All Neutrino Interaction
Generator (ANIS)~\cite{bib:ANIS}. ANIS incorporates a model of the earth 
interior
with the appropriate density changes between inner and outer core and mantle.
It offers two alternatives for the extrapolation of neutrino cross sections 
on nucleons:
A smooth power-law extrapolation of pQCD CTEQ5 structure functions, and a hard
pomeron enhanced extrapolation. The two differ by about a factor of
three at 10$^{20}$~eV.
Our calculations were done with the lower cross section extrapolation
that did not have the pomeron enhancement.
As far as the cross sections are concerned our limit should therefore be
conservative, but neutrino cross section extrapolation is the major
systematic uncertainty in this analysis.

ANIS also incorporates code for the propagation and decay of the tau leptons.
The intrinsically stochastic energy losses at high energy are approximated by
a smooth energy loss function.
For HiRes the ensuing suppression of potential sub-showers resulting from
catastrophic energy loss events along the path of the tau lepton is not much
of a concern, as the trigger threshold of the detector even for very close
events is above $10^{16}$~eV.
Tau decays in ANIS are modeled with the TAUOLA package.

As ANIS was designed for underground detectors, it has neither an atmosphere
nor any topographic structure on the earth surface built into it.
We integrated both a US standard desert atmosphere~\cite{bib:atm} and the
detailed surface topography of the surroundings of the HiRes
detectors. The elevation model we incorporated is
based on a 30~arcsec grid~\cite{bib:topo}.

\begin{figure}[h]
\begin{center}
\includegraphics [width=0.98\textwidth]{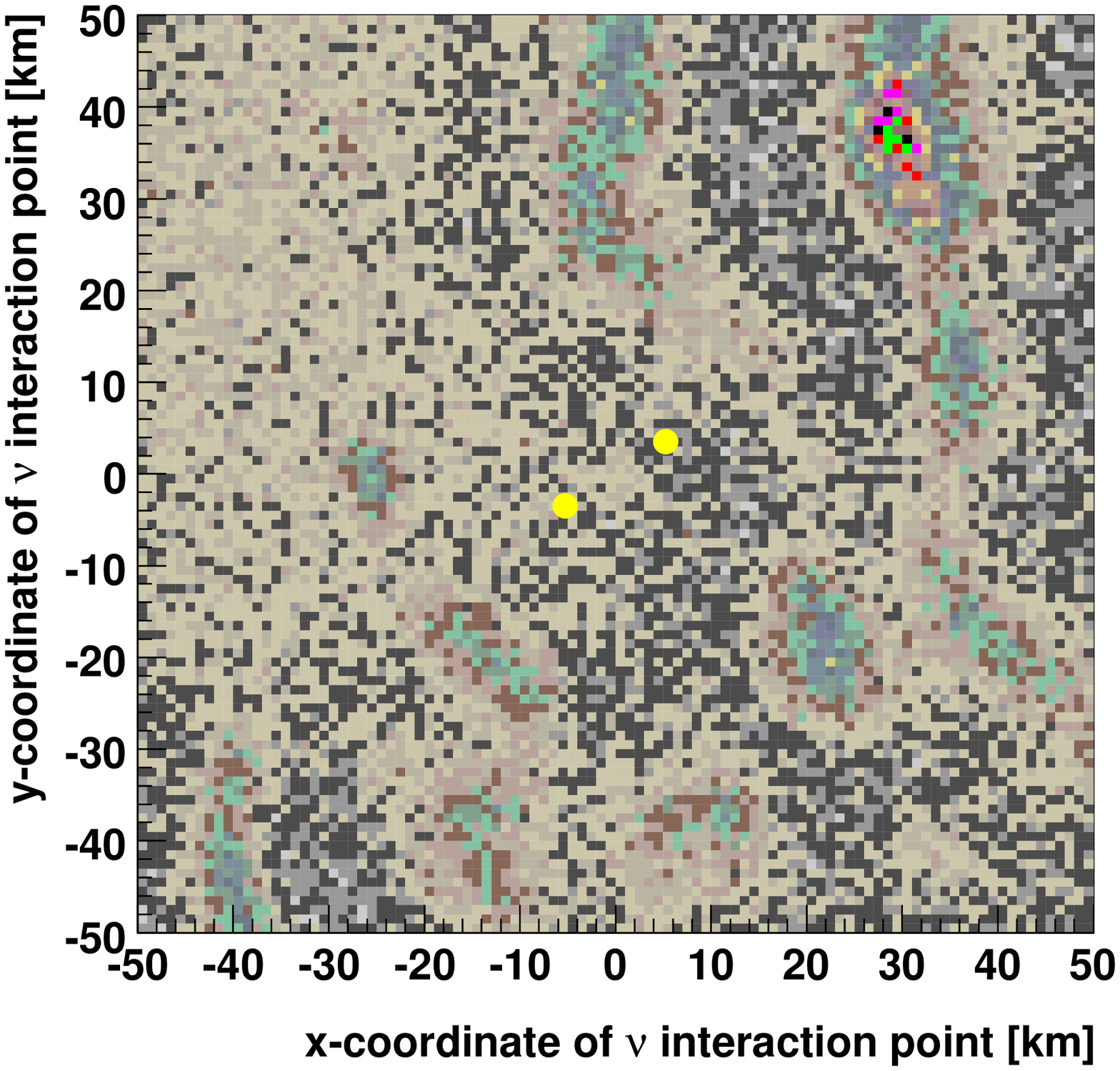}
\end{center}
\caption{A map of the surroundings of the HiRes detectors made from 
the interaction points of tau neutrinos for which the tau leptons decay in the
atmosphere. The mountains are visible in green, and two yellow dots mark 
the positions of the HiRes detectors.}
\label{fig:topo}
\end{figure}

Figure \ref{fig:topo} is a scatter plot of the $\nu_\tau$ interaction points 
for $\nu_\tau$ where the ensuing tau lepton decays above ground. 
The figure clearly highlights the role of the surrounding mountains as
target mass.
It can also be seen that HR1 has greater proximity to the bulk of that target
mass.
While $\nu_\tau$ interactions in the atmosphere are not suppressed in this
simulation, they do not play a role due to the limited target mass available.

\begin{figure}[h]
\begin{center}
\includegraphics [width=0.98\textwidth]{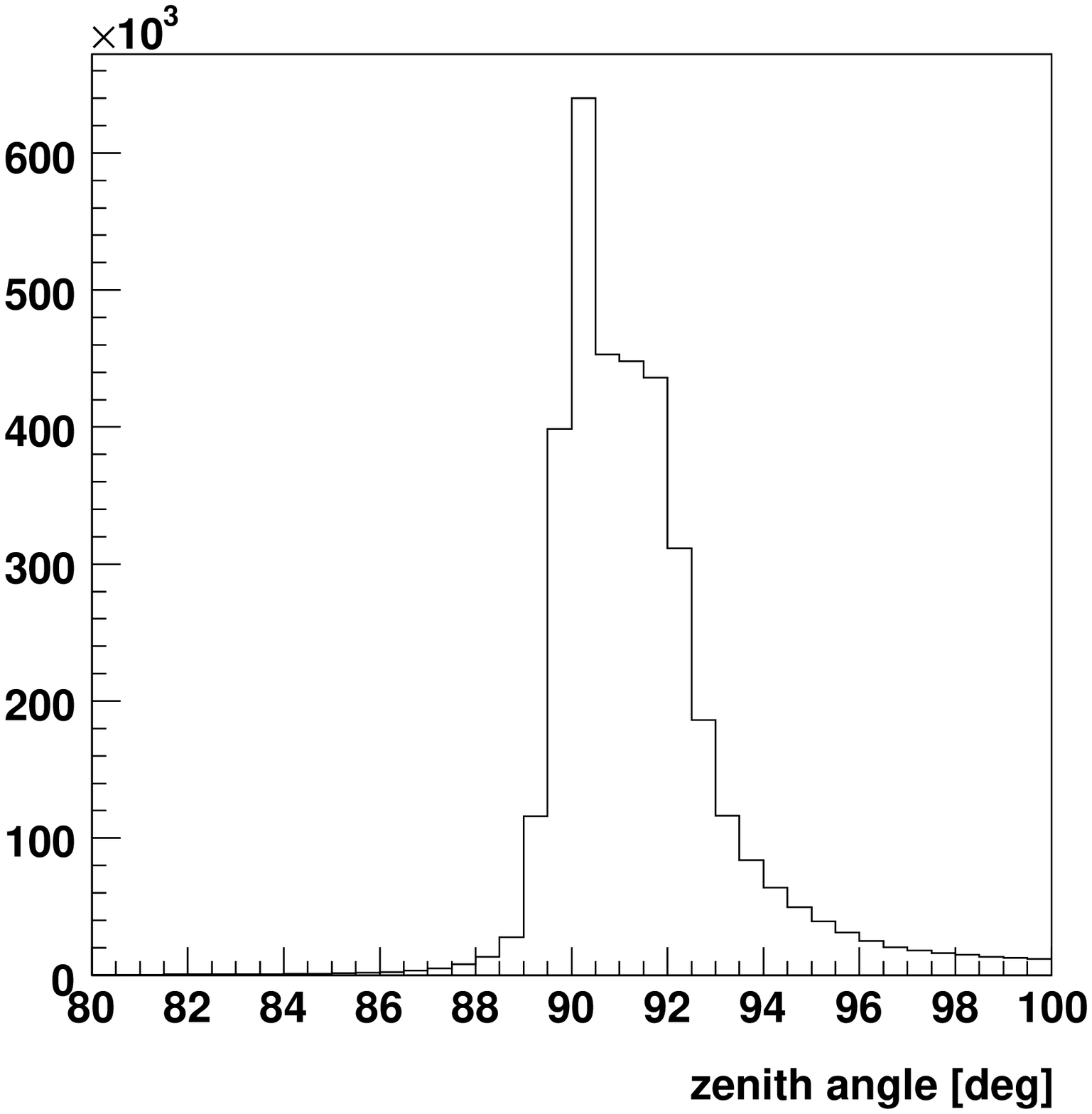}
\end{center}
\caption{Zenith angles of tau lepton directions for tau leptons that decay
in the atmosphere.}
\label{fig:zenith}
\end{figure}

A prominent feature of the tau leptons that emerge into the atmosphere
is that their zenith angles are concentrated within a few degrees from
horizontal.
Figure \ref{fig:zenith} shows the distribution of zenith angles of the
tau leptons that decay in the atmosphere.
Zenith angles larger than 90~degree are upward going.
We exploit this feature to constrain the
neutrino injection directions to $\pm$10~degree above and below the local
horizon, thus saving significant computing time in our simulations.
We also limit the impact parameter of neutrino trajectories to within 75~km
of the geometrical center between HR1 and HR2.

The tau leptons decay to 70\% into hadronic decay channels. Charged pion
initiated showers are used in the subsequent air shower simulation.
In case a leptonic decay channel is entered, the ensuing lower energy
tau is followed up upon, a muon is given up as unobservable, and an electron
is fed into the appropriate electromagnetic shower simulation.

For the hadronic decays air shower simulation is carried out with the help of
CORSIKA \cite{bib:CORSIKA}
version 6.200. Shower libraries are generated using QGSJET \cite{bib:QGSJET}
that contain 400
quasi-horizontal hadronic showers each in a matrix of thirteen evenly spaced
energies
from 10$^{15}$~eV to 10$^{21}$~eV and nine evenly spaced heights between 1~km
and 5~km above ground.
After a suitable tau decay is identified in the ANIS output, a set of shower
parameters is chosen randomly from the appropriate library, and scaled from
the nearest energy found in the library to the energy extracted from the ANIS
simulation.
These shower parameters and the ANIS generated event location and direction
are then handed over to the standard version of
the HiRes stereo Detector Monte Carlo (DMC) program.
If the tau lepton decays to an electron, the DMC itself generates the
appropriate profile for an electromagnetic shower.

The DMC generates fluorescence (and Cherenkov) light according to the
geometry and energy of the shower, propagates both through the atmosphere
towards the detector, and performs a detailed simulation of the detector
response.

494~M tau neutrinos with impact parameters up to 75~km were injected into
ANIS.
Their input energies were distributed
between $10^{18}$ and $10^{21}$~eV according to an $E^{-2}$ differential
spectrum.
While there was no constraint on the azimuth angle of the neutrino injection,
the zenith angle was constrained to within $\pm$10~degree of the horizon.
From this input we got a total of
5829 events that triggered the detector: 4243 triggers in HR1, and 2456 in
HR2. 870 of these events triggered both detectors. 

Our electron neutrino simulation uses the same cross section extrapolation as 
the tau neutrino analysis. Using a spherically symmetric model of the earth 
with just one transition between the mantle with density 4.6~g/cm$^3$ and the 
crust with 2.8~g/cm$^3$ at 58.4~km below the surface the MC simulation is much 
more efficient as no terrain features break the symmetry. The LPM effect is 
implemented following the formalism of \cite{bib:lpm} to calculate the 
energy dependence for the probabilities to undergo bremsstrahlung or pair 
production. 

\section{HiRes Data and Analysis}

The HR1 data used in this analysis stem from the equivalent of 20,132,360
seconds of operation with all HR1 mirrors and were taken between 05/1997 and
11/2005. The HR2 data were taken over the equivalent of 13,096,693 seconds of
operation with all HR2 mirrors between 10/1999 and 11/2005.
This data set includes 10,128,727 seconds of stereo operation. A total of 
75 million data records were analyzed. In a first step noise events and 
artificial light sources are eliminated. 

The relevant variables characterizing the EAS geometry are the zenith angle
$\Theta$ and the impact parameter $R_p$ of the 
shower axis with respect to the relevant detector location. 


Two defining characteristics identify our $\nu_\tau$ events: \\
- They are horizontal (zenith angle)\\
- They are low in the atmosphere ($R_p$ distance) \\
The last criterion is not trivial as very high up in the atmosphere CR events 
can develop almost horizontally. 
In order to reliably extract $\nu_\tau$ events we impose cuts on the quality 
of the event reconstruction. 
After these quality cuts and application of an $R_p < 20$~km cut events with 
reconstructed zenith
angles between 88.8 and 95.1 degree (1.55 and 1.66 radians)
are kept as neutrino candidates.
With these cuts we keep 366 neutrino MC events in HR1 and 209 in HR2,
out of which 21 are stereo events.

Applying the same criteria to the data yields 75 events from HR1 and 59 events 
from HR2. All of these 134 neutrino candidate events are unfortunately laser 
events. The reason that these events pass all our laser cuts have to do with 
weather: Light scattered off haze near the ground lets these particular 
events reconstruct as horizontal. Unlike true neutrino events their individual 
geometries can all be shown to repeat that of another event inside or outside 
of the set of selected neutrino candidates. Our conclusion is that HiRes has 
zero tau neutrino candidates. 

Data analysis for the electron neutrino search is based on the superior 
reconstruction afforded by the HR2 detector's flash ADC system, and therefore 
is restricted to events seen with the HR2 detector. As in the tau neutrino 
analysis no neutrino candidate events were found. 

\begin{table}
\begin{center}
\begin{tabular}{|c||c|c|c|}
\hline
log$_{10}$(E/eV): & 18 - 19 & 19 - 20 & 20 - 21 \\
\hline
\hline
$\nu_\tau$ & 420$^{+25}_{-20}$ & 1340$^{+140}_{-110}$ & 29400$^{+22100}_{-8900}$ \\
$\nu_e$ & 3820$\pm 120$ & 3260$\pm 80$ & 4250$\pm 100$ \\
\hline
\end{tabular}
\caption{Fluxes excluded by HiRes at the 90\% CL in units of
eV~sr$^{-1}$s$^{-1}$cm$^{-2}$ for three energy ranges. The uncertainties
derive from MC statistics only. \label{tab:results}}
\end{center}
\end{table}

\begin{figure}[h]
\begin{center}
\includegraphics [width=0.98\textwidth]{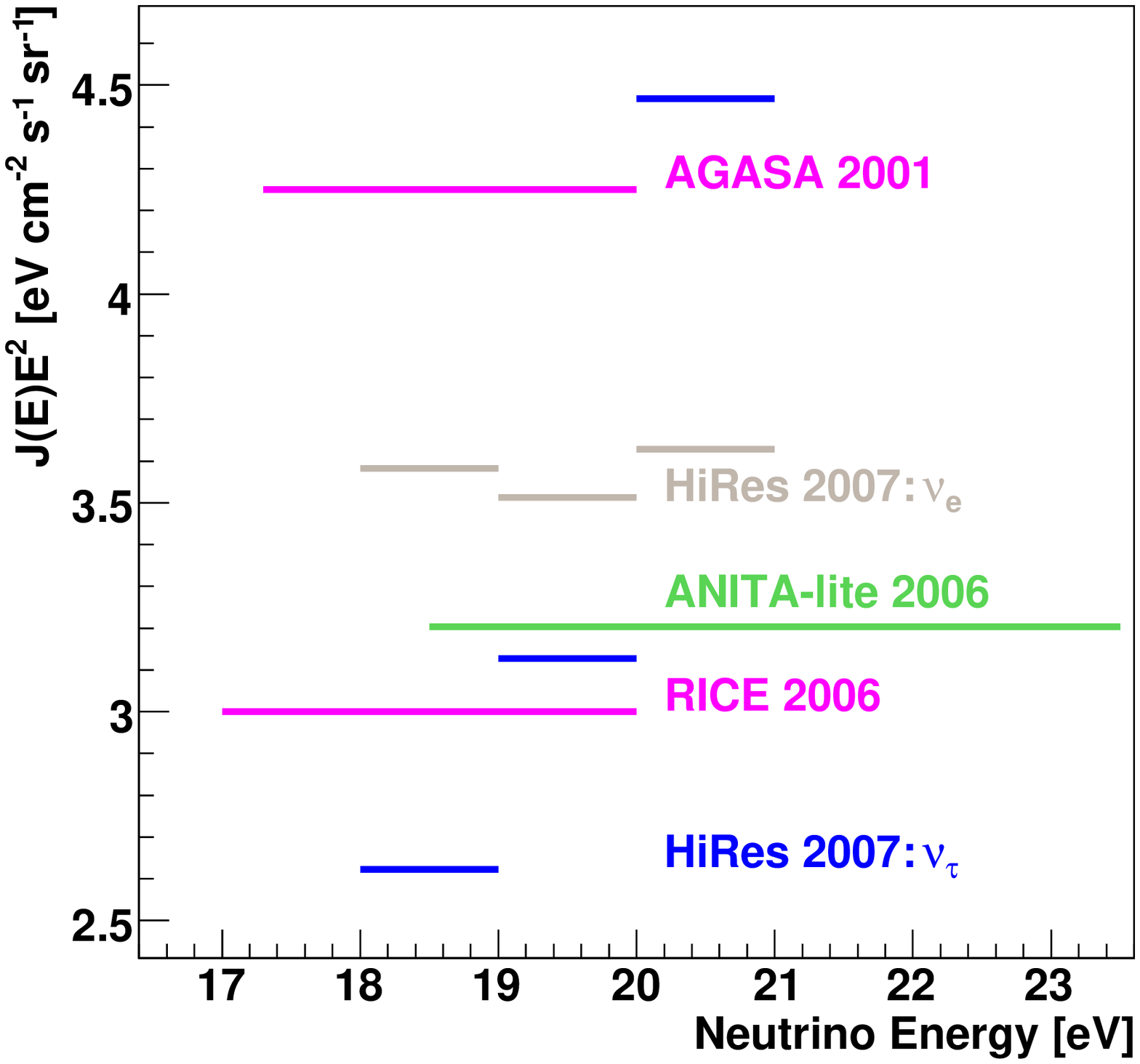}
\end{center}
\caption{The HiRes neutrino flux limits in context.}
\label{fig:limits}
\end{figure}

Our limits on neutrino fluxes are based on Feldman/Cousins unified confidence 
intervals and are listed in table \ref{tab:results} as well as shown in the 
context of other experimental results in figure \ref{fig:limits}.

\section{Conclusions}

Searching for near horizontal showers HiRes has set new 
competitive limits on cosmogenic neutrino fluxes. 
At 10$^{19}$~eV 
the tau lepton decay length (not counting energy losses) reaches 500~km, 
so that despite the growing cross section for neutrino interaction the 
subsequent tau lepton decay at higher energies increasingly is pushed beyond 
the physical limits of the atmosphere. LPM delayed electron neutrino showers 
do not have this handicap, leading to a competitive result at the highest 
energies. All of these limits are orders of magnitude above the flux 
expectation we derived from our cosmic ray spectrum. 

\section{Acknowledgements}

This work is supported by US NSF grants PHY-9100221, PHY-9321949,
PHY-9322298, PHY-9904048, PHY-9974537, PHY-0073057, PHY-0098826,
PHY-0140688, PHY-0245428, PHY-0305516, PHY-0307098, PHY-0649681, and
PHY-0703893, and by the DOE grant FG03-92ER40732.  We gratefully
acknowledge the contributions from the technical staffs of our home
institutions and the Utah Center for High Performance Computing. 
The cooperation of Colonels E.~Fischer, G.~Harter and
G.~Olsen, the US Army, and the Dugway Proving Ground staff is greatly
appreciated.

\bibliographystyle{plain}

\begin{thebibliography}{}
\bibitem[1]{bib:GZK} K. Greisen, PRL 16 (1966) 748
  G.T. Zatsepin and V.A. Kuzmin, Pisma Zh. Experim. Theor. Phys. 4 (1966) 114
\bibitem[2]{bib:HR-GZK} HiRes Collaboration, astro-ph/0703099
\bibitem[3]{bib:Comp} R. Abbasi et al., astro-ph/0501317 Phys. Lett. B 619 (2005), 271
\bibitem[4]{bib:Doug} D.R. Bergman, Nucl.Phys.Proc.Suppl.136:40-45, 2004 
\bibitem[5]{bib:Berezinsky} V. Berezinsky, A. Gazisow, and S. Grigorieva, PRD 74 (2006), 043005
\bibitem[6]{bib:ANIS} ANIS: High Energy Neutrino Generator for
Neutrino Telescopes, astro-ph/0406439
\bibitem[7]{bib:atm} D.R. Longtin, Air Force Geophysics Laboratory
,
AFL-TR-88-0112 (1988)
\bibitem[8]{bib:topo} http://edc.usgs.gov/products/elevation/gtopo30/
gtopo30.html
\bibitem[9]{bib:CORSIKA} D. Heck et al., Report FZKA 6019 (1998)
\bibitem[10]{bib:QGSJET} N.N. Kalmykov, S.S. Ostapchenko, and
A.I. Pavlov, Nucl. Phys. B (Proc. Suppl.) 52B (1997) 17
\bibitem[11]{bib:lpm} T. Stanev et al., Phys. Rev. D 25 (1982), 1291
\end{thebibliography}

\end{document}